\documentclass[fleqn,12pt]{wlscirep}
\usepackage[utf8]{inputenc}
\usepackage[T1]{fontenc}
\usepackage{lineno}
\usepackage{comment}
\usepackage{subcaption}
\usepackage{multirow}
\nolinenumbers

\title{Gap-free 16-year (2005-2020) sub-diurnal surface meteorological observations across Florida }

\author[1,*]{Julie Peeling}
\author[1]{Jasmeet Judge}
\author[2]{Vasubandhu Misra}
\author[2]{C. B. Jayasankar}
\author[3]{Rick Lusher}
\affil[1]{University of Florida, Center for Remote Sensing, Department of Agricultural and Biological Engineering, Gainesville, FL 32611, USA}
\affil[2]{Florida State University, Center for Ocean-Atmospheric Prediction Studies, Department of Earth, Ocean and Atmospheric Science, Tallahassee, FL 32306, USA}
\affil[3]{University of Florida IFAS Extension, Florida Automated Weather Network, Gainesville, FL 32611, USA}

\affil[*]{corresponding author(s): Julie Peeling (juliepeeling@ufl.edu)}

\begin{abstract} 
The rather unique sub-tropical, flat, peninsular region of Florida is subject to a unique climate with extreme weather events across the year that impacts agriculture, public health, and management of natural resources. Meteorological data at high temporal resolutions especially in the tropical latitudes are essential to understand diurnal and semi-diurnal variations of climate, which are considered to be the fundamental modes of climate variations of our Earth system. However, many meteorological datasets contain gaps that limit their use for validation of models and further detailed observational analysis. The objective of this paper is to apply a set of data gap filling strategies to develop a gap-free dataset with 15-minute observations for the sub-tropical region of Florida. Using data from the Florida Automated Weather Network (FAWN), methods of linear interpolation, trend continuation, reference to external sources, and nearest station substitution were applied to fill in the data gaps depending on the extent of the gap. The outcome of this study provides continuous, publicly accessible surface meteorological observations for 30 FAWN stations at 15-minute intervals for the years 2005-2020. 
\end{abstract}
\begin{document}
\flushbottom
\maketitle
\begin{sloppypar}

\thispagestyle{empty}

\section*{Background \& Summary}
Data such as rainfall, temperature, wind patterns, and solar radiation are significant meteorological variables in determining climate variations and change. For example, high spatial and temporal resolution rainfall data is necessary for the development of hydrological models, flood risk assessment, land management, and climate model validation \cite{huang_hourly_2022, ochoa-tocachi_high-resolution_2018, takhellambam_temporal_2022, dai_diurnal_2023}. In sub-tropical, flat regions such as Florida, slight seasonal climate shifts can have drastic impacts on flooding, agricultural production, and public health \cite{dewan_developing_2022, Raimi_Florida_2020, misra_florida_2021, misra_impacts_2020}. Florida is a sub-tropical region with average air temperatures fairly stable in the summer across the state and varying from North to South (increasingly warm) in the winter \cite{misra_impacts_2020}. During summer months, average temperatures are typically between 24 and 28 $^\circ$C (~297-301 K) across the state, and in winter, Northern Florida averages around 7-13 $^\circ$C (280-286 K) while the southern part of the state tends to average around 15-19 $^\circ$C (288-292 K) \cite{Shin_FLclimate_2020}. The elevation levels in Florida ranges from sea level to about 105 m above sea level \cite{SFRC_Elevation_2016}. Due to its peninsular geography in subtropical latitudes and interactions with relatively warm oceans, Florida has a unique climate to the rest of the United States \cite{misra_characterizing_2017, misra_operational_2022}. Its wet season is heavily interconnected with fresh water availability and ecosystem functionality, and as population growth continues throughout the state, there is the further strain placed on its natural resources \cite{misra_characterizing_2017}.

\par Within the climate system, the diurnal and semi-diurnal scale variations represent a fundamental mode of variability \cite{yang_diurnal_2001}. Diurnal variations are generated from diurnally varying solar heating that affects near the surface, through the depths of the troposphere, and in the stratosphere that manifests as pronounced oscillations with periods of approximately 24 h (diurnal) and 12 h (semi-diurnal). These periodic oscillations that appear in the upper atmosphere are also called atmospheric tides, which significantly impact the diurnal and semi-diurnal variations of many climatic variables \cite{Chapman_Lindzen_1970}. Often, the fidelity of numerical climate and weather models and reanalyses is assessed in the ability to represent the diurnal scales owing to its feature of being a fundamental mode of variation of the climate system (e.g., \cite{Lin_2000, Trenberth_2003, Dai_Trenberth_2004, Lee_2007}). However, a huge limitation of verifying these models to simulate the diurnal cycles is the lack of data that robustly resolves the diurnal variations \cite{Dai_2006}. To examine these variations in sub-tropical areas such as the state of Florida, continuous data is needed at sub-diurnal (hourly or finer) temporal and spatial resolution.
\par The Florida climate is representative of a trade wind regime for latitudes between about 25 degrees North and South of the equator, including monsoon regions such as India and Vietnam \cite{misra_characterizing_2017, misra_understanding_2017,vu_uncertainties_2018}. Areas such as these often lack a high-density observational network, but for the Florida region, many climate and weather datasets exist to provide information at varying spatio-temporal resolutions. For example, the Florida Climate Center\renewcommand{\thefootnote}{\Roman{footnote}} \footnote{The Florida Climate Center is located at Florida State University's Center for Ocean-Atmospheric Prediction Studies and is part of a three-tiered system providing climate services at the national, regional and state levels.} provides daily precipitation and temperatures for approximately 100 stations across Florida, and hourly local climatological data (LCD) is available through the National Oceanic and Atmospheric Administration (NOAA) \cite{NOAA_LCD_2021,mukherjee_climate_2017}. These datasets lack a sub-hourly temporal resolution, limiting their applicability. In addition, there are two precipitation-only data sources. Integrated Multi-Satellite Retrievals for Global Precipitation Mission (IMERG) data from the National Aeronautics and Space Administration (NASA) provides 30-minute precipitation data at a 10 km resolution over the period June 2000-present \cite{misra_operational_2022}. The second source, from the National Climatic Data Center (NCDC), offers 15-minute precipitation observations from stations that are sparsely located (31 stations throughout Florida). 
 \cite{NCDC_15min_PPT}. 
\par
\begin{figure}[ht]
\centering
\includegraphics[width=0.8\linewidth]{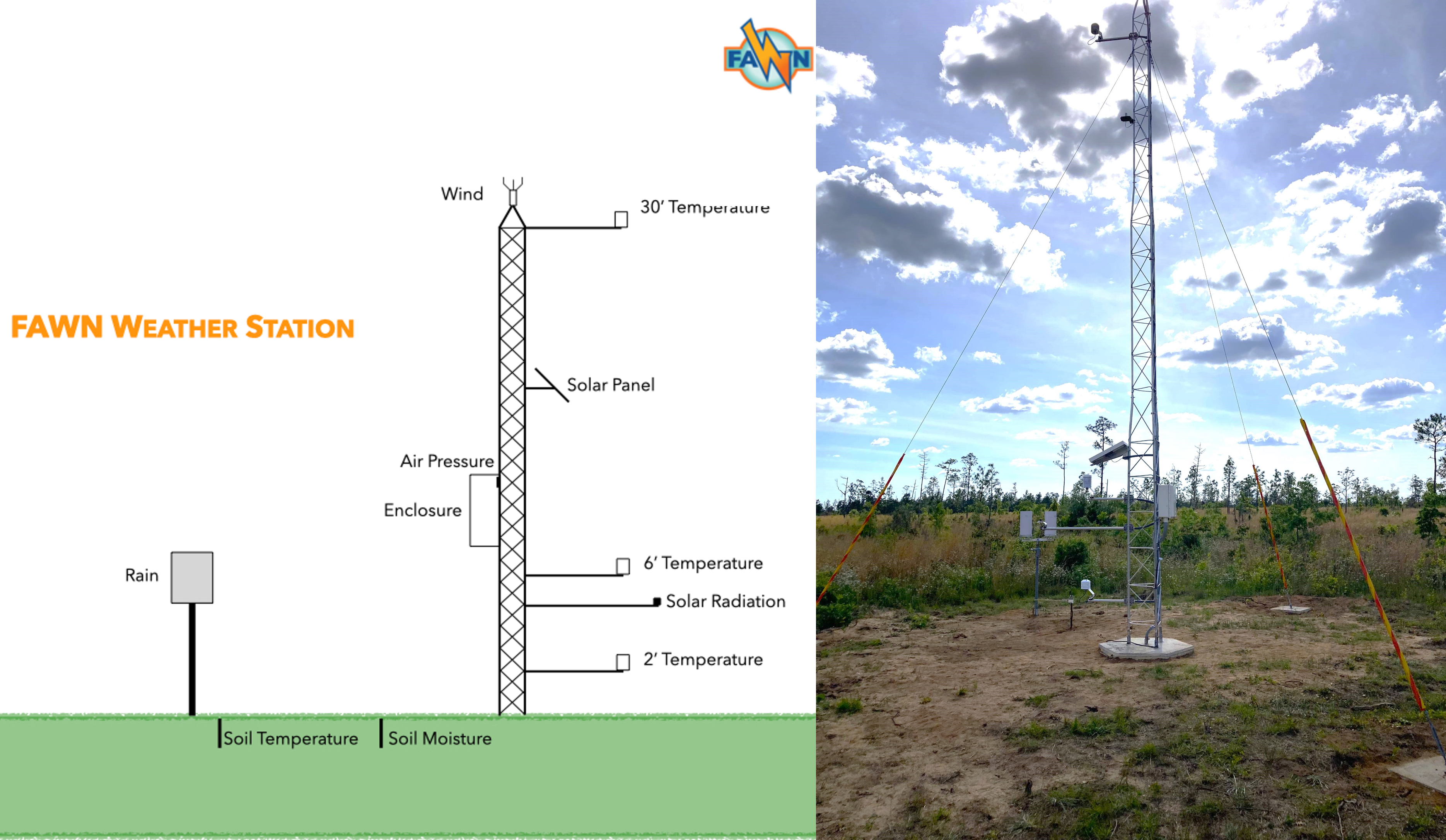}
\caption{Typical FAWN tower configuration (left) and photo of a FAWN tower (right).}
\label{fig:FAWNtower}
\end{figure}

 The Florida Automated Weather Network (FAWN) is the only network providing sub-hourly data for 10 meteorological variables \cite{FAWN_2023}. Initiated in 1997 to provide climatic data to rural areas in Florida to inform growers, it is currently comprised of 42 stations  \cite{Lusher_2008_article}. The goal of FAWN is to provide accurate, reliable, and real-time weather data to users across Florida for applications such as cold weather protection strategies, irrigation scheduling, and extreme precipitation analysis \cite{Lusher_2008_article,Jackson_2008, morgan_use_2011,zhang_interpreting_2022}. Each FAWN tower, as shown in Figure \ref{fig:FAWNtower}, is equipped with sensors that measure air temperature at 60 cm (T 60cm), air temperature at 2 m (T 2m), air temperature at 10 m (T 10m), soil temperature (T Soil), relative humidity (RH), precipitation (PPT), wind speed (WS), wind direction (WD), solar radiation (Sol Rad), and barometric pressure \cite{Lusher_2008_article}. The FAWN dataset also includes derived parameters such as dew point temperature (T Dew), wet bulb temperature, and potential evapotranspiration \cite{Lusher_2008_article}. However, barometric pressure, wet bulb temperature, and potential evapotranspiration are not included in annual datasets. FAWN data is offered at a fine temporal resolution of 15 minutes, but many of these datasets contain gaps of various sizes, from 15 minutes to one year, due to operational issues that limit their use for applications that require continuous datasets \cite{Lusher_2008_article}. To address this challenge, this paper details the methods utilized to fill gaps in FAWN meteorological observations to generate a continuous dataset over 2005-2020. Previous studies have employed various data homogenization and gap-filling methods for meteorological variables, such as wind speed and precipitation, across Florida to improve prediction methods and better understand trends in extreme weather \cite{huang_neural_2009, peng_coastal_1999, zhang_interpreting_2022}. The value of the the dataset generated in this study lies in its fine temporal resolution and diverse set of meteorological variables. This study leverages the FAWN infrastructure in order to create a gap-free dataset for wider scientific applications in regions of similar characteristics. The publicly available dataset will provide a unique resource within a complex sub-tropical region for climate analysis and modeling.

\section*{Methods}
\subsection*{Data Acquisition and Preprocessing}
\begin{figure}[ht]
\centering
\includegraphics[width=\linewidth]{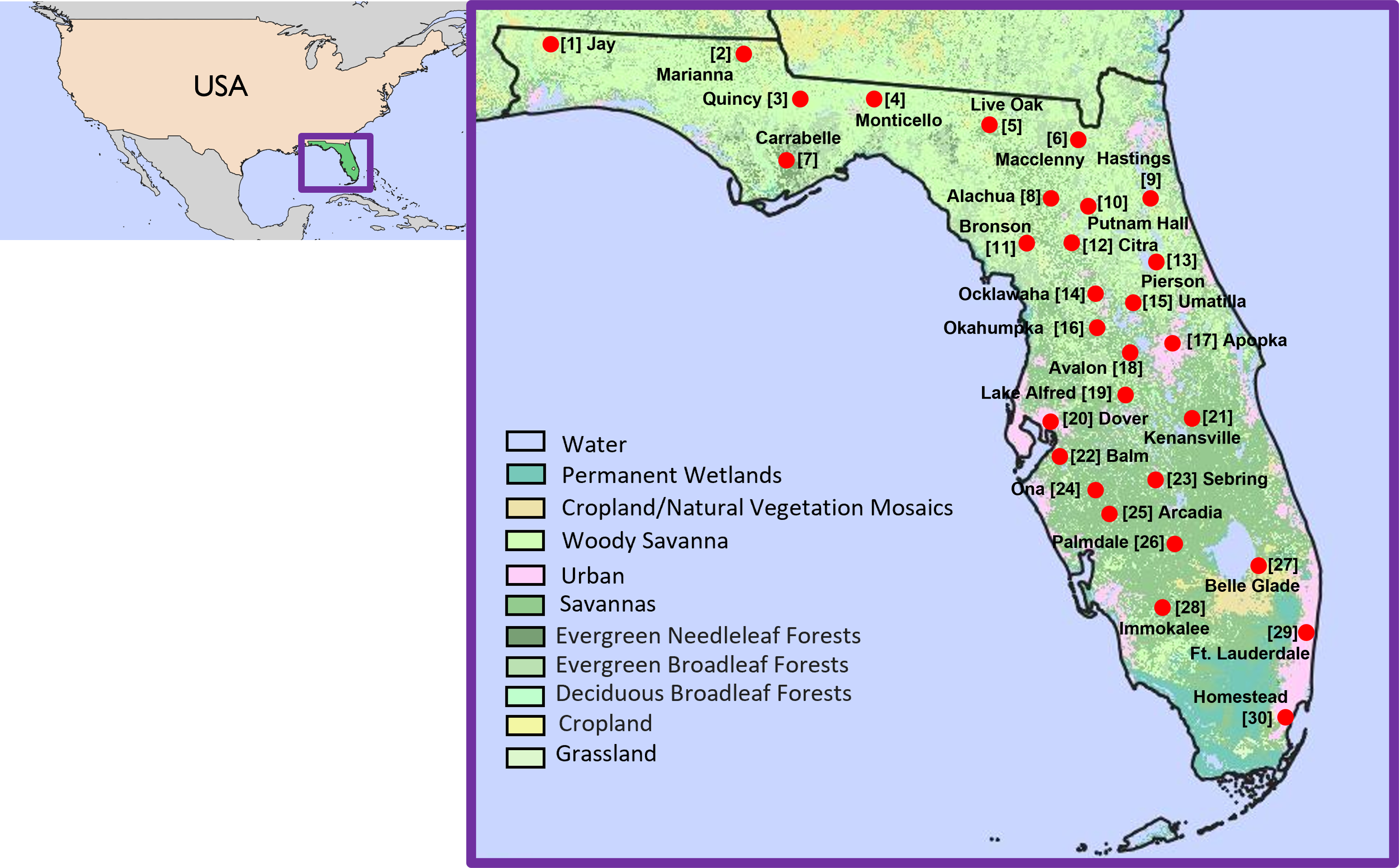}
\caption{Location of 30 FAWN stations selected for this study across Florida, along with their names and numbers. The base map is a 2019 Land Cover map from Moderate Resolution Imaging Spectroradiometer (MODIS).}
\label{fig:FLmap}
\end{figure}

Yearly observations at 15-minute intervals were obtained from FAWN for all active stations \cite{FAWN_2023}. In this study, we examined the FAWN data available between 1997 and 2020 and selected the stations with data present across the longest period of time during which the most stations were available, resulting in the chosen 30 stations over 2005-2020, as shown in Figure \ref{fig:FLmap}. In the northern part of the State, 16 stations were located in forested and woody environments, and in the South, nine stations were in areas classified as savanna. Four of the stations were positioned in urban areas, and one station was located in cropland.
\par FAWN implements initial quality control measures and filtering before publishing the raw data, details for which are given in Table \ref{tab:QC}. Annual tests are conducted to determine if repair or replacement of sensors is needed based upon EPA guidelines, and filtering of these potential operational incidents as well as power failures result in data gaps \cite{EPA_QualityGuide_2000}. In this study, supplemental quality control mechanisms were implemented to enhance the data reliability. For all temperature measurements, if there was a difference > 5 $^{\circ}$ C within one time step of 15 minutes, the data point was marked as a data gap that was filled as described below. For WS higher than 30 mph, the event was manually checked against nearby FAWN stations and LCD reports to confirm high wind speeds. If the high wind speeds were confirmed, then they remained in the data set, and if they could not be verified, the value was marked as a data gap. Additionally, RH values of 0\% were marked as data gaps.
\subsection*{Gap Filling}
\begin{figure}[ht]
\centering
\includegraphics[width=\linewidth]{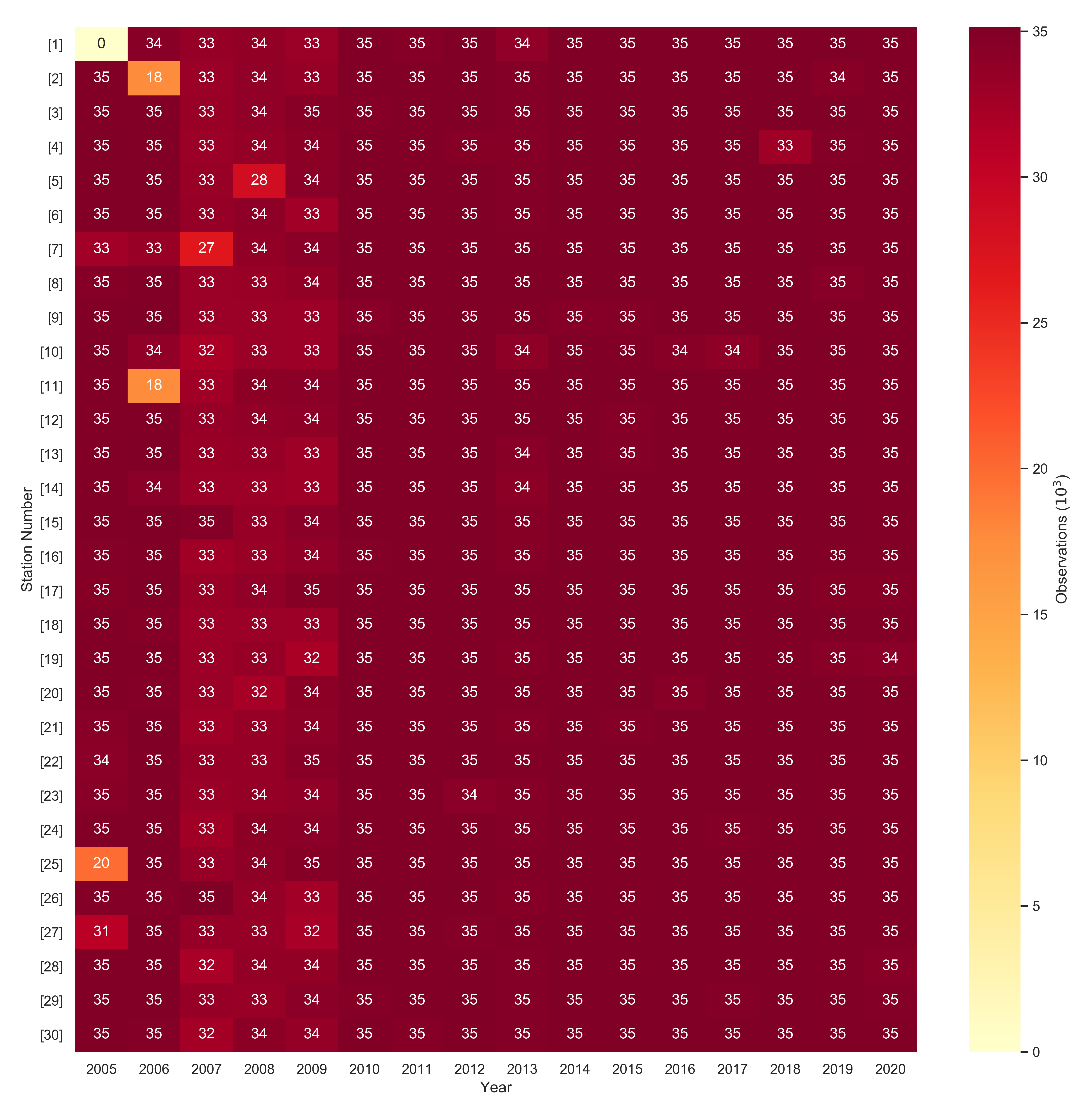}
\caption{Heat map representing the number of observations (in thousands) present in the raw data for each station in each year.}
\label{fig:ObsHeat}
\end{figure}

Data gaps occurred if the difference between two consecutive data points was greater than 15 minutes. The number of 15-minute observations in the raw data for each station at each year is given in Figure \ref{fig:ObsHeat} to provide insight into the amount of data points present and the extent of missing data. Figures \ref{fig:MinMax2}a and \ref{fig:MinMax2}b provide the minimum and maximum number of consecutive 15-minute data gaps missing for each station in each year, respectively, to demonstrate the distribution of data gap extents across space and time. Station \#s 7 and 8 had a larger number of 15-minute gaps than other stations, with the most gaps occurring in 2007 and 2008, respectively. The years 2007-2009 had the most gaps for all stations, with over 1000 gaps for most stations during that period. Large data gaps such as these are primarily due to operational issues, generally from power failures. Gap filling of meteorological variables is inherently uncertain and challenging, with differing methodological approaches for different variables \cite{Longman2018DataIslands, Richard_2021, Luedeling_2022, denhard_evaluation_2021, henn_comparison_2013, Graf_2017}. We applied several methods of data gap filling based upon gap size and the nature of the meteorological variables. 

\par
\begin{figure}[ht]
\centering
\includegraphics[width=\linewidth]{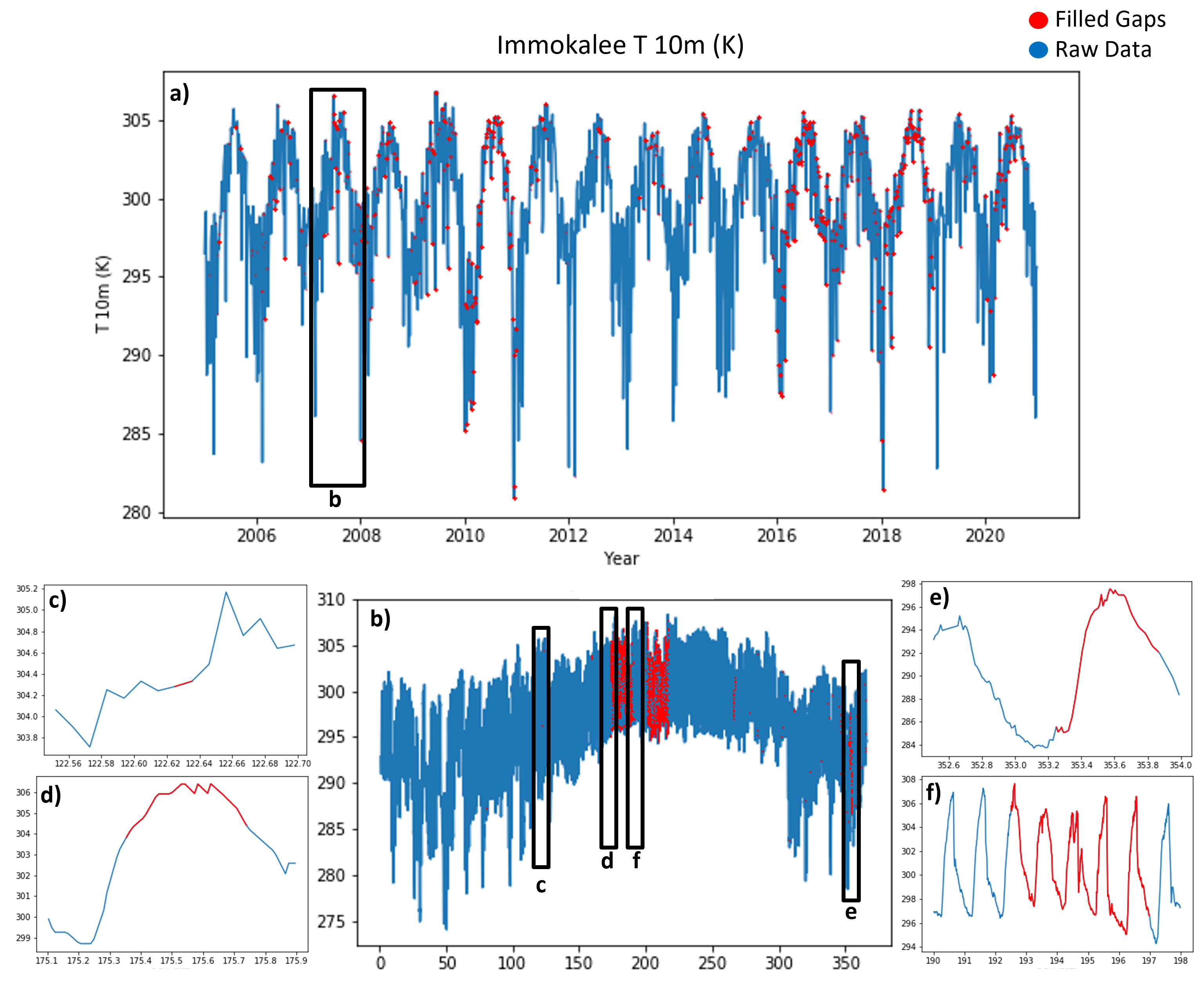}
\caption{Examples for the methods used for filling diurnal meteorological variable such as 10 m air temperature at the Immokalee station (\#28): a) 16-year time-series data; b) Zoomed in times-series for 2007; c) linear interpolation for gaps less than 6 hours; d) trend continuation method for gaps between 6-12 hours; and e) external data source and f) nearest station for gaps larger than 12 hours.}
\label{fig:methods}
\end{figure}

 \textbf{Datasets with Diurnal Cycles}. Gap filling for datasets with diurnal cycles such as  temperatures, RH, and Sol Rad followed the same methodology. Figure \ref{fig:methods}a depicts an example of gap filling for T 10m over the study period for station \#28. The year 2007 had the most gaps for this station (see Figure \ref{fig:methods}b), and various gap filling techniques were applied based on the gap size. For data gaps < 6 hours (about 82\% of gaps), \textit{linear interpolation} was implemented using the slope between the two data points at either end of the gap to estimate the missing data points (see Figure \ref{fig:methods}c) \cite{Luedeling_2022, denhard_evaluation_2021, henn_comparison_2013}. Such temporal interpolation is a reliable data gap filling method in continuous climate variables such as near-surface air temperature and solar radiation data \cite{Luedeling_2022, denhard_evaluation_2021, henn_comparison_2013}. For gaps between 6 and 12 hours (about 1\% of gaps), \textit{trend continuation} for the meteorological variables was implemented, similar to Tardivo \cite{Tardivo_2012} and Kemp \cite{Kemp_1983}, by extracting the data values and trends from two days prior to and two days after gaps (see Figure \ref{fig:methods}d). At each missing time step, the measurement was filled with the average values at that particular time from the surrounding days. When the data gaps were greater than 12 hours (about 17 \% of gaps), an outside data source was referenced. These large gaps mainly occurred in the years 2005-2009 for most stations (see Figures \ref{fig:ObsHeat} and \ref{fig:MinMax2}). In this study, LCD from NOAA \cite{NOAA_LCD_2021} was used as an \textit{external data source} to fill these large data gaps with data from weather stations within the same city, or if not available then the same county (see Figure \ref{fig:methods}e). Hourly LCD values were linearly interpolated to 15-minute intervals. Since LCD is available only for T 10m, T Dew, and RH, so larger data gaps in T 60cm, T 2m, T Soil and Sol Rad were filled using data from the nearest FAWN station (see Figure \ref{fig:methods}f)\cite{Richard_2021, Luedeling_2022, Graf_2017}. The nearest station was determined through the smallest euclidean distance to a station with available data. The temporal correlations were high between the monthly means of these nearest stations. The nearest station method was also applied for any periods when there were gaps in the LCD, following Luedeling \cite{Luedeling_2022} and Graf \cite{Graf_2017}. In this study, the distance to the nearest station was typically around 32 km. 
\par \textbf{Discrete Datasets}. To fill gaps in the discrete datasets such as PPT, WS, and WD, LCD and nearest FAWN stations were used, similar to the larger data gaps mentioned above \cite{Richard_2021, Luedeling_2022, Graf_2017}. The NCDC PPT data could not be used due to large distances from FAWN stations and lack of observations consistent with the FAWN and LCD PPT observations. Given the distribution of available FAWN stations, it was reasonable to assume that gradients of observed data for these meteorological variables were captured by filling the gaps using the nearest station.

\section*{Data Records}
Gap-free data for 30 FAWN stations over the period 2005-2020 are available through Figshare, an open access repository, in CSV file format titled "\textit{Gap-Free Sub-Diurnal Meteorological Data from Florida Automated Weather Network (FAWN)}". The data is continuous over 16 years for each station listed in Figure \ref{fig:FLmap}, and annual data within the given time period can be downloaded. There are 10 gap-filled meteorological variables provided in the datasets, the units and labels of which are given in Table \ref{tab:dataRecord}. 

\section*{Technical Validation}
\par In addition to visual inspection of filled data such as comparing diurnal patterns with surrounding days, the validity of the data was assessed to ensure consistency between the filled data and the raw data for each station and meteorological variable. This was assessed by conducting differential statistics between the raw data and the filled data. A two-tailed T-test on the means of each meteorological variable at each station was conducted to determine whether the mean of the filled data differed significantly from the mean of the raw data \cite{henn_comparison_2013}. This test was chosen as one source of validation in order to ensure that the gap filling process did not significantly alter the mean of the filled data as compared to the raw data. All p-values resulting from the T-test were > 0.1, so there was no significant difference found between the filled data means and raw data means (see Table \ref{tab:StatTests} for minimum p-values). 
\par Figures \ref{fig:heatMap2}a and \ref{fig:heatMap2}b provide the mean, along with the standard deviation, minimum, and maximum values, for the 10 meteorological variables at each station. As expected, the mean air temperature values increase from station \#1, at around 292 K, to station \#30, at around 297 K (North to South). The maximum PPT was highest, between 52.1 mm and 68.3 mm, at station \#s 3, 8, 21, and 28, providing information on the areas which received the highest intensity rainfall within a 15-minute period over the study period. The standard deviation of the temperature values tended to decrease from station 1 (around 8 K) to 30 (around 5 K), supporting higher temperature variability in the more northern stations. 
\par To test the difference in standard deviations between the meteorological variables at each station in the filled dataset and raw dataset, an F-test was implemented. As we determined that the means of the filled and raw data were not significantly different, this test was conducted to reveal whether the dispersal of value around the averages of each dataset significantly varied. These p-values resulting from the F-test were also all > 0.1, indicating no notable difference in standard deviations. 
\par In order to test the statistical difference in distribution of the raw and filled datasets, the Kolmogorov-Smirnov (KS) test was implemented. This test essentially checks whether two datasets come from the same distribution, and the test statistic can be interpreted to represent the greatest distance between the cumulative distribution function of each dataset \cite{Nguyen_2013}. Thus, the KS test was chosen as a third validation metric to determine if there existed significant difference between the shape and spread of the filled and raw datasets. The resulting p-values from the KS test showed no such difference, as they were all > 0.1.

\section*{Usage Notes}

The gap-filled dataset generated through this work is unmatched in temporal resolution and spatial extent across the state of Florida. It provides information on 10 meteorological variables at 15-minute intervals, spanning 30 stations from as far north as Jay (latitude 31 $^\circ$ N) to Homestead in the south (latitude 25.5 $^\circ$ N). It also has potential applications in climate monitoring, agriculture, and hydrology. The gap free data can be applied to understand climate variability and verify numerical climate and weather models, which can be used to predict future weather conditions from current observation \cite{mitra_use_2023}. The continuous 16-year data product developed through the methods outlined above can serve as an important resource for climate research and forecasting in sub-tropical regions such as Florida \cite{Dai_2006, misra_characterizing_2017}.

\section*{Code Availability}
No custom software was used to process the data described in this paper. The open-source software used to conduct this study was Python version 3.7.6. The packages and libraries used included Numpy (V 1.18.1), Pandas (V 1.3.0), Matplotlib (V 3.1.3), and Scipy (V 1.4.1). Specific functions for the statistical analysis including the T-test, F-test, and Kolmogorov-Smirnov functions were conducted using the Scipy stats module.

\bibliography{FAWNbib}

\section*{Acknowledgements}

This study was supported by a grant from NASA Award \#80NSSC19K1199.

\section*{Author contributions statement}

J.J., V.M., and J.P. were responsible for the conceptualization; R.L. and J.P. were responsible for data handling and quality control; J.P., J.J., V.M., and J.B. developed the methodology; J.B., J.J., V.M., and J.P. contributed to visualization and writing (the original draft);  All authors contributed to reviewing and editing. All authors have read and agreed to the published version of the manuscript.

\section*{Competing interests}
The authors declare that they have no known competing financial interests or personal relationships that influenced the work reported in this paper. 




\begin{table}[ht]
\small
\centering
\begin{tabular}{|l|r|r|}
\hline
\textbf{Meteorological Variable} & \textbf{Minimum} & \textbf{Maximum} \\
\hline
All Temperatures ($^{\circ}$C) & -20 & 50 \\
\hline
Relative Humidity (\%) & 0 & 100 \\
\hline
Precipitation (in) & 0 & 3 \\
\hline
Solar Radiation (W/m$^{2}$) & 0 & 1200 \\
\hline
Wind Speed (mph) & 0 & 75 \\
\hline
Wind Direction ($^{\circ}$) & 0 & 360 \\
\hline
\end{tabular}
\caption{\label{tab:QC} Quality control measures implemented by FAWN, adapted from the FAWN measurement system specifications \cite{FAWNQC_2023}. }
\end{table}

\begin{figure}[ht]
\centering
    \begin{subfigure}{\textwidth}
        \includegraphics[width=\linewidth]{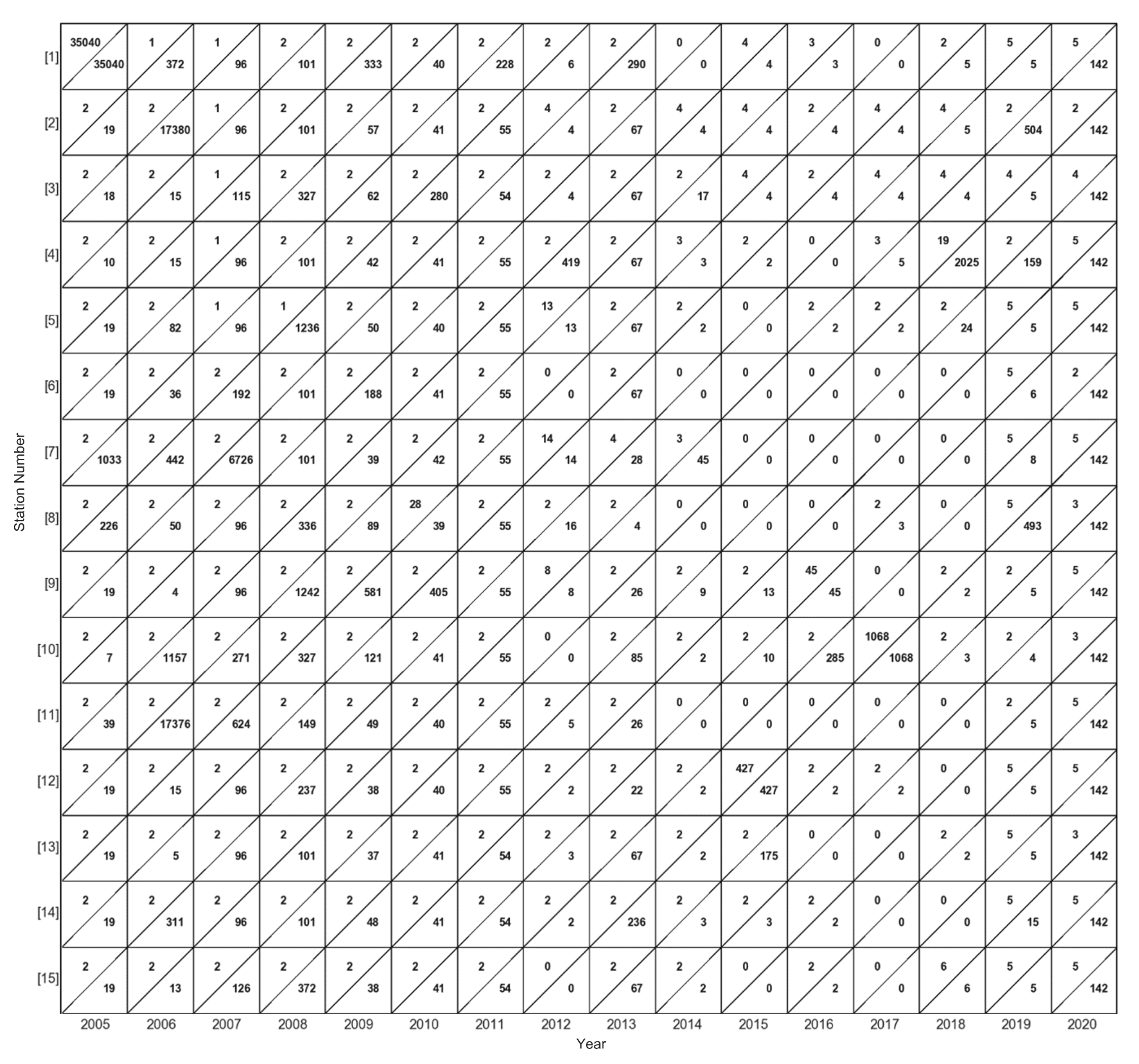}
        \subcaption{}
        \label{fig:4a}
    \end{subfigure}
\label{fig:MinMax1}
\end{figure}

\begin{figure}[ht]\ContinuedFloat
\centering
    \begin{subfigure}{\textwidth}
        \includegraphics[width=\linewidth]{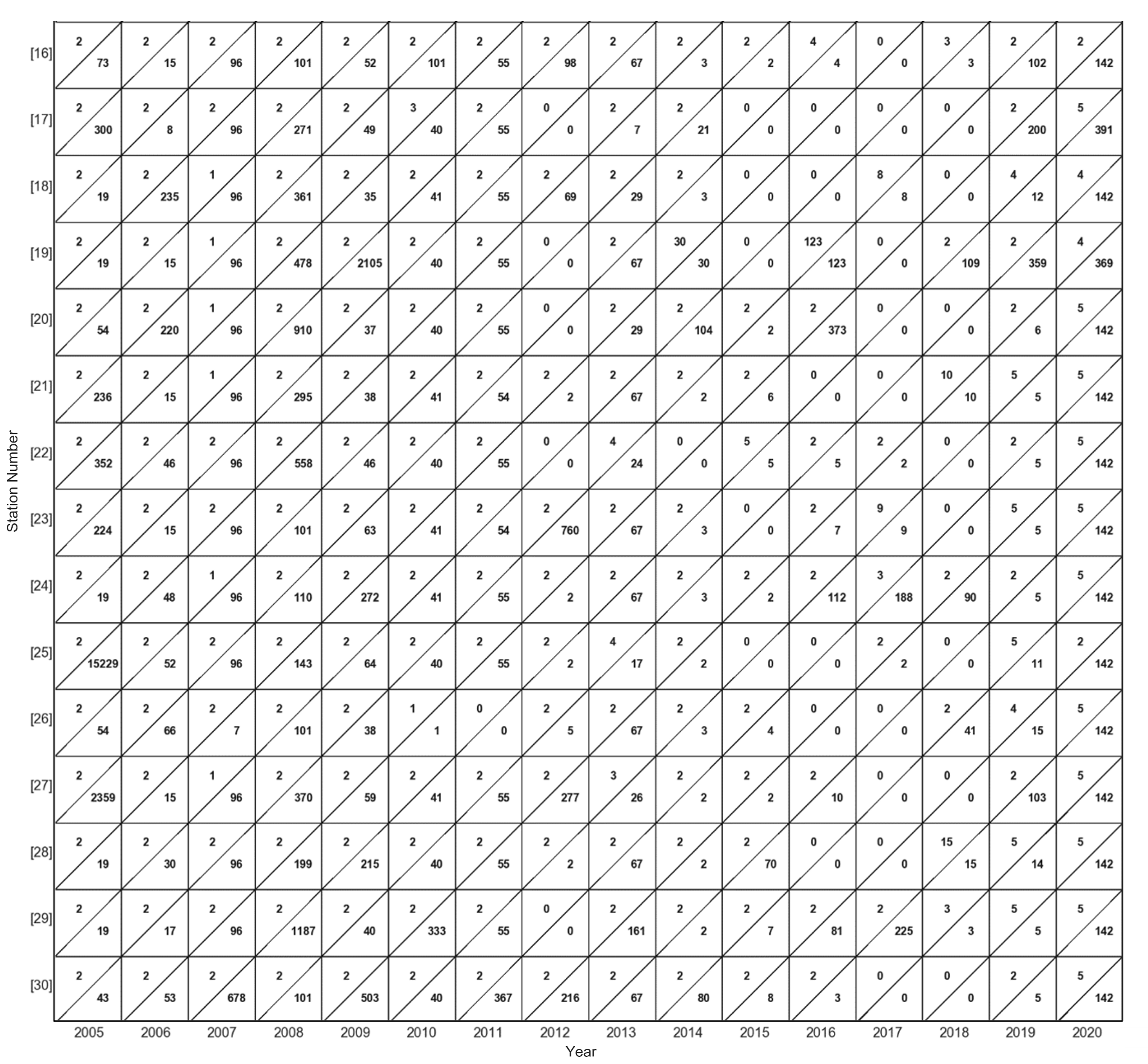}
        \subcaption{}
        \label{fig:4b}
    \end{subfigure}
\caption{The minimum (left triangle) and maximum (right triangle) number of 15-minute data gaps present at each station in each year for a) station \#s 1-15 and b) station \#s 16-30.}
\label{fig:MinMax2}
\end{figure}

\begin{table}[ht]
\small
\centering
\begin{tabular}{|l|l|}
\hline
\multicolumn{1}{|c|}{\textbf{Meteorological Variable}} & \multicolumn{1}{c|}{\textbf{Description}}\\ 
\hline
T 60cm & Air Temperature at 60 cm ($^{\circ}C$)\\ 
T 2m & Air Temperature at 2 m ($^{\circ}C$)\\ 
T 10m & Air Temperature at 10 m ($^{\circ}C$)\\ 
T Dew & Dew Point Temperature ($^{\circ}C$)\\ 
T Soil & Soil Temperature ($^{\circ}C$)\\
RH & Relative Humidity (\%)\\
PPT & Rainfall Amount (inches) \\
WS & Wind Speed (mph) \\
WD & Wind Direction (degrees) \\
Sol Rad & Solar Radiation (W/m$^{2}$) \\
\hline
\end{tabular}
\caption{\label{tab:dataRecord} Micrometeorological variables and their descriptions as available from FAWN on an annual basis for download.}
\end{table}

\begin{table}[ht]
\small
\centering
\begin{tabular}{|l|c|c|c|}
\hline
\textbf{Meteorological Variable} & \textbf{T-test p-value} & \textbf{F-test p-value} & \textbf{KS p-value} \\
\hline
T 60cm & 0.104 & 0.483 & 0.149 \\
\hline
T 2m & 0.216 & 0.466 & 0.289 \\
\hline
T 10m & 0.274 & 0.438 & 0.374 \\
\hline
T Dew & 0.549 & 0.291 & 0.595 \\
\hline
T Soil & 0.818 & 0.460 & 0.855 \\
\hline
RH & 0.374 & 0.343 & 0.375 \\
\hline
PPT & 0.812 & 0.500 & 0.815 \\
\hline
WS & 0.683 & 0.389 & 0.700 \\
\hline
WD & 0.705 & 0.385 & 0.693 \\
\hline
Sol Rad & 0.750 & 0.496 & 0.750 \\
\hline
\end{tabular}
\caption{\label{tab:StatTests}Minimum p-values for each meteorological variable from the statistical significance tests, including the T-test, F-test, and Kolmogorov-Smirnov (KS) test, on each station.} 
\end{table}

\begin{figure}[ht]
\centering
    \begin{subfigure}{\textwidth}
        \includegraphics[width=\linewidth]{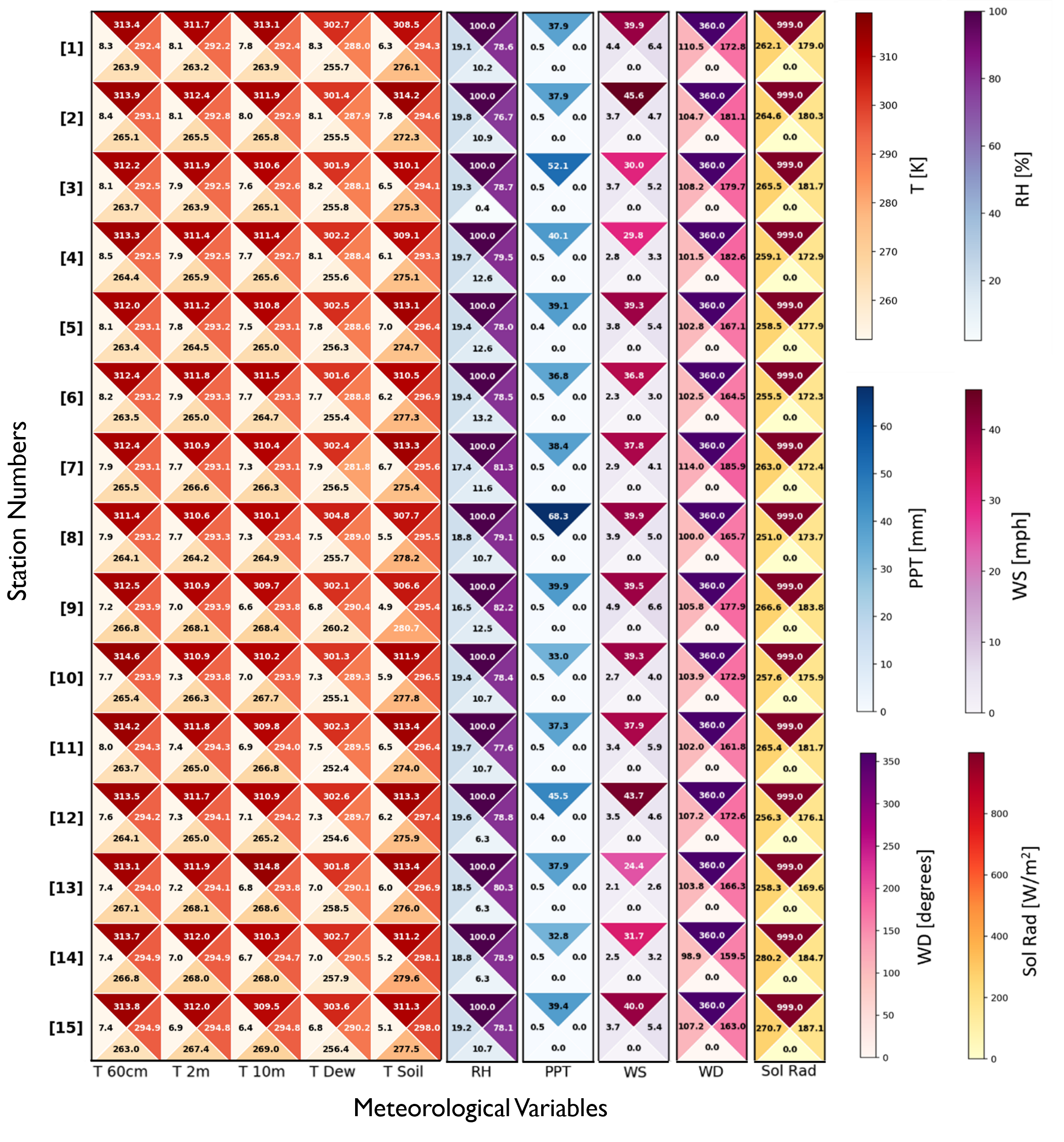}
        \subcaption{}
        \label{fig:6a}
    \end{subfigure}
\label{fig:heatMap1}
\end{figure}

\begin{figure}[ht]\ContinuedFloat
\centering
    \begin{subfigure}{\textwidth}
        \includegraphics[width=\linewidth]{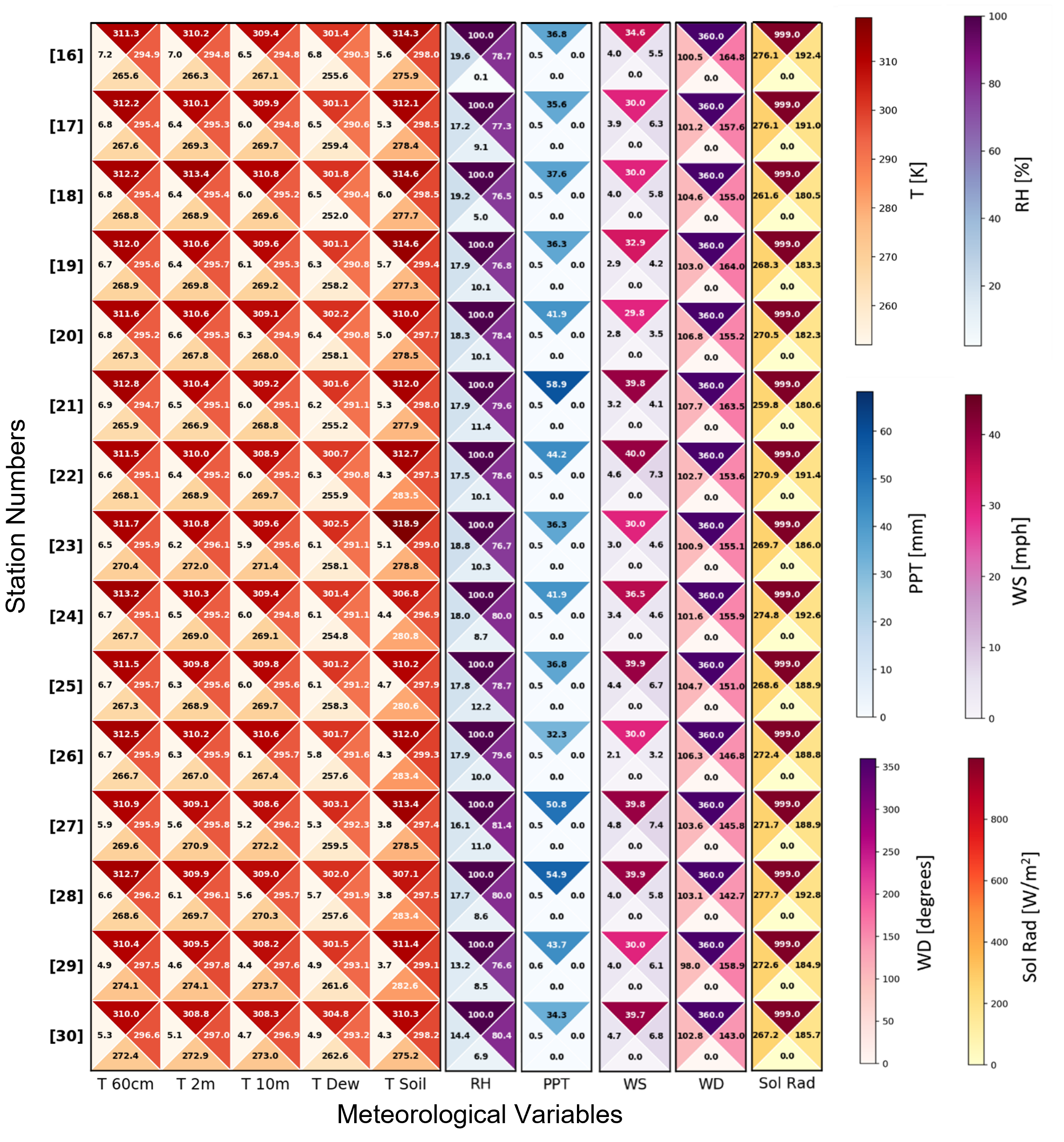}
        \subcaption{}
        \label{fig:6b}
    \end{subfigure}
\caption{Statistical description for the 10 meteorological variables at (a) station \#s 1-15 and (b) station \#s 16-30, provided in four triangles. In the clockwise direction from the top, each triangle provides the maximum, mean, minimum, and standard deviation of the gap-filled dataset over the 16-year period.}
\label{fig:heatMap2}
\end{figure}
\end{sloppypar}

\end{document}